\documentclass[conference,a4paper]{IEEEtran}

\pdfpageattr{/Group << /S /Transparency /I true /CS /DeviceRGB >>}

\usepackage[utf8]{inputenc}
\usepackage[T1]{fontenc}
\usepackage{silence}
\usepackage{textcomp}
\usepackage{fixltx2e}
\usepackage{microtype}
\usepackage{calc}
\usepackage{xspace}
\usepackage{xcolor,listings}
\usepackage{textcomp}
\usepackage{lipsum}  

\usepackage[forbid-literal-units,range-phrase=--,per-mode=symbol-or-fraction,binary-units=true,range-units=single,list-units=single,detect-all]{siunitx}
\usepackage{silence}
\AtBeginDocument{
\DeclareSIUnit\byte{Byte}
\DeclareSIUnit\decibeli{dBi}
\DeclareSIUnit\decibelm{dBm}
\DeclareSIUnit\megasamplespersecond{Msps}
\DeclareSIUnit\dbm{dBm}
\DeclareSIUnit\ppm{ppm}
\DeclareSIUnit\watthour{Wh}
\sisetup{output-exponent-marker=\textsc{e}}
}

\usepackage{csquotes}
\usepackage[backend=biber,style=ieee,doi=false,isbn=false,maxnames=99,mincitenames=1,maxcitenames=2]{biblatex}
\DeclareFieldFormat{sentencecase}{#1} 
\DeclareFieldFormat{titlecase}{#1} 
\usepackage{xpatch}
\xpatchbibmacro{textcite}{\addspace}{\addnbspace}{}{}
\xpatchbibmacro{Textcite}{\addspace}{\addnbspace}{}{}
\DefineBibliographyStrings{english}{
	andothers = et~al\adddot\addspace
}

\DeclareSourcemap{
	\maps[datatype=bibtex]{
		\map[overwrite=true]{
			\step[fieldsource=school,fieldtarget=institution]
		}
	}
}

\usepackage{amsmath}
\usepackage{amssymb}
\usepackage{amsfonts}
\usepackage{amsthm}

\usepackage{booktabs}
\usepackage{url}

\usepackage[caption=false,font=footnotesize]{subfig}
\usepackage[pdftex]{graphicx}
\DeclareGraphicsExtensions{.pdf,.png,.jpg}
\usepackage{tikz}
\usetikzlibrary{calc}

\usepackage[american]{babel}
\usepackage{hyphenat}

\usepackage[capitalize,noabbrev,nameinlink]{cleveref}

\RequirePackage{xstring}
\RequirePackage{xparse}
\RequirePackage[]{acro}
\makeatletter
\@ifpackagelater{acro}{2019/02/17}{
	\NewDocumentCommand\acrodef{mO{#1}mG{}}{\DeclareAcronym{#1}{short={#2}, long={#3}, foreign-plural={}, #4}}
}{
	\NewDocumentCommand\acrodef{mO{#1}mG{}}{\DeclareAcronym{#1}{short={#2}, long={#3}, #4}}
}
\makeatother

\acrodef{ap}[AP]{access point}
\acrodef{RF}{radio frequency}
\acrodef{VLC}{Visible Light Communication}
\acrodef{LoS}{line-of-sight}
\acrodef{NLoS}{non-line-of-sight}
\acrodef{MIMO}{Multiple-Input Multiple-Output}
\acrodef{COTS}{commercially available off-the-shelf}
\acrodef{CSI}{channel state information}
\acrodef{IoT}{internet of things}
\acrodef{LPWAN}{low-power wide area network}

\acrodef{BB}{broadband}
\acrodef{TS}{time-sensitive}
\acrodef{DL}{downlink}
\acrodef{UL}{uplink}
\acrodef{MTAP}{Multi-Technology Access Point}
\acrodef{CTC}{Cross-Technology Communication}
\acrodef{FCS}{Frame Check Sequence}
\acrodef{CCK}{Complementary Code Keying}
\acrodef{HOL}{Head-of-Line}
\acrodef{IFS}{interframe space}
\acrodef{BO}{backoff}
\acrodef{SDR}{Software-defined Radio}
\acrodef{CIR}{Channel Impulse Response}
\acrodef{LOS}{Line-of-sight}
\acrodef{CFR}{Channel Frequency Response}
\acrodef{AP}{Access Point}
\acrodef{STA}{Station}
\acrodef{TTL}{time to live}

\usepackage{todonotes}
\usepackage[inline]{enumitem}

\newcommand{\proposalName}{\texttt{Ns3Sionna}}

\begin{document}


\title{Ns3 meets Sionna: Using Realistic Channels in Network Simulation}


\author{
\IEEEauthorblockN{
Anatolij Zubow,
Yannik Pilz,
Sascha Rösler,
Falko Dressler
}
\IEEEauthorblockA{
    School of Electrical Engineering and Computer Science, TU Berlin, Germany
}%
\texttt{\{zubow,pilz,roesler,dressler\}@tkn.tu-berlin.de}
}

\maketitle 
\begin{abstract}
Network simulators are indispensable tools for the advancement of wireless network technologies, offering a cost-effective and controlled environment to simulate real-world network behavior. 
However, traditional simulators, such as the widely used ns-3, exhibit limitations in accurately modeling indoor and outdoor scenarios due to their reliance on simplified statistical and stochastic channel propagation models, which often fail to accurately capture physical phenomena like multipath signal propagation and shadowing by obstacles in the line-of-sight path. 
We present \proposalName{}, which integrates a ray tracing-based channel model, implemented using the Sionna RT framework, within the ns-3 network simulator. 
It allows to simulate environment-specific and physically accurate channel realizations for a given 3D scene and wireless device positions.
Additionally, a mobility model based on ray tracing was developed to accurately represent device movements within the simulated 3D space. 
\proposalName{} provides more realistic path and delay loss estimates for both indoor and outdoor environments than existing ns-3 propagation models, particularly in terms of spatial and temporal correlation. 
Moreover, fine-grained channel state information is provided, which could be used for the development of sensing applications.
Due to the significant computational demands of ray tracing, \proposalName{} takes advantage of the parallel execution capabilities of modern GPUs and multi-core CPUs by incorporating intelligent pre-caching mechanisms that leverage the channel's coherence time to optimize runtime performance.
This enables the efficient simulation of scenarios with a small to medium number of mobile nodes.

%
%
%
\end{abstract}

\begin{IEEEkeywords}
Ns3, Network Simulator, Sionna, channel model, ray tracing
\end{IEEEkeywords}

\maketitle

\acresetall

\section{Introduction}\label{sec:intro}
Wireless communication has become increasingly vital, establishing itself as an integral component of modern life. 
Wireless technologies like IEEE 802.11 Wi-Fi are ubiquitous, present in nearly every household, workplace, and numerous public spaces. 
The development of new wireless technologies and protocols relies heavily on rigorous experimentation for testing and deployment. 
However, conducting real-world experiments is often costly and time-intensive, requiring physical hardware and controlled environments.
Consequently, there is significant interest in simulation tools capable of replicating wireless networks and accurately modeling communication within them. Such tools enable researchers to validate and evaluate emerging technologies and protocols in a cost-effective, scalable, and reproducible manner.

Several network simulators are available, including OPNET, OMNeT++, and ns-3. 
Among these, ns-3 stands out as an open-source platform that supports a wide range of communication protocols. 
It offers superior performance in terms of runtime and memory efficiency compared to other simulators\,\cite{weingartner2009performance}. 
As a result, ns-3 has gained widespread popularity in both industry and academia as a reliable tool for studying communication technologies and network protocols\,\cite{zubow2023toward}.

Despite their advantages over real-world experiments, network simulators have limitations in reproducing the characteristics and dynamics of a wireless channel\,\cite{yoo2014toward}.
To simulate physical phenomena such as signal propagation, these tools rely on statistical and stochastic models.
While these models are applicable for certain scenarios, they particularly lack realism in complex environments like indoor spaces~\cite{dricot2004high}, where signals interact with various objects like walls and furniture.
The statistical and stochastic models provided by network simulators do not account for strong indoor multipath or model effects such as fading or shadowing based on probability distributions.
Therefore, their use for simulating wireless propagation limits the ability to accurately predict the behavior of real networks~\cite{wilhelmi2021usage}.

An alternative to statistical and stochastic models is ray tracing, a technique that models radio waves as rays~\cite{tseng2017ray-tracing-assisted} and predicts their propagation in an environment~\cite{yun2015raytracing}.
Ray tracing can calculate the reflection, diffraction, and scattering of radio waves, enabling it to accurately estimate the propagation channel between a transmitter and receiver~\cite{tseng2017ray-tracing-assisted}, including multipath propagation.
Additionally, ray tracing allows the computation of spatially consistent \acp{CIR}, which statistical and stochastic models in network simulators cannot achieve~\cite{hoydis2023learning}.
These characteristics are necessary for numerous research topics in the field of wireless communication~\cite{hoydis2023sionnart} and the development of wireless systems~\cite{degli-esposti2014ray-tracing}.

The open-source library Sionna\,\cite{hoydis2023sionna} includes a ray tracing extension that enables the precise simulation of radio wave propagation in user-defined 3D environments.
Sionna allows modeling of both indoor and outdoor environments, considering the different materials of scene objects and their electromagnetic properties.
This capability makes it possible to accurately calculate the reflection, diffraction, and scattering of radio waves in an environment in order to replicate the wireless channel.

\smallskip

\noindent \textbf{Contribution:} We present \proposalName{}, a software module which integrates ray tracing based propagation modeling using Sionna into the ns-3 network simulator with the aim of enabling a very accurate simulation of the wireless channel, particularly in indoor environments, with a focus on scenarios using 802.11 Wi-Fi technology.
%
%
Comparative analyses reveal that \proposalName{} provides more realistic path and delay loss estimates for both indoor and outdoor environments than existing statistical and stochastic propagation models in ns-3, particularly in terms of spatial and temporal correlation. 
We address the very high computational demands of ray tracing by incorporating an intelligent pre-caching which leverages the channel's coherence time to optimize run-time performance.
The advantage of the parallel execution capabilities of modern GPUs and multi-core CPUs is utilized by avoiding the calculation of each point-to-point (P2P) channel individually and instead to offload the computation of the entire point-to-multipoint (P2MP) channel, i.e. from the sender to all potential receivers, to Sionna.
Moreover, pre-caching is realized by also calculating the channels which might be required in the future.


\section{Background}\label{background}
%

%
%
\subsection{Wireless Channel Propagation}\label{sec:channel}
A radio signal propagating through the wireless channel to the receiver over multiple paths experiences several effects.
Reflection on walls or scattering through obstacles will produce additional copies of the transmitted signal,
so-called multipath components, which have different arrival times.
This behavior of the channel can be seen by different impulses in the \ac{CIR}.
When assuming a time-invariant channel, the \ac{CIR} can be denoted as~\cite{yang2013from}:
\begin{equation}
	h(t) = \sum_{n=0}^{N} a_n e^{-j\phi_n} \delta(t-t_n)
\end{equation}
where $a_n$ is the amplitude, $\phi_n$ is the phase of the $n$th multipath component at the time $t$, $N$ is the total number of multipath components and $\delta(t)$ is the Dirac delta function.
The sum of amplitude and phase of these multipath components can result in constructive or destructive interference which is referred to as small-scale fading~\cite{rappaport2009wireless}.
An important parameter that describes the stability of the wireless channel over time is the coherence time $T_C$. 
It describes the period of time over which the wireless channel remains relatively stable and where there is a high correlation between the signal amplitudes~\cite{rappaport2009wireless}. 
If the coherence time is defined as the time period during which the time correlation of the channel is greater than 0.5, it can be calculated using the following formula~\cite{rappaport2009wireless}:

\begin{equation}\label{eq:coherence_time_fundamentals}
    T_C = \frac{9}{16\pi}\frac{c}{v \cdot f}
\end{equation}

where $c$ is the speed of light, $f$ the transmission frequency and $v$ the relative velocity between the transmitter and receiver.
If the relative speed is high, the channel therefore changes faster and thus the coherence time is shorter than in static scenarios in which the transmitter and receiver are stationary.

%
%
%
%

Another effect that can influence signal reception is shadowing. 
In contrast to fading, which describes short-term amplitude fluctuations due to multipath propagation, shadowing leads to fluctuations in signal strength over a period of seconds or minutes and occurs when larger objects such as buildings block the signal path between the transmitter and receiver~\cite{tse2005fundamentals}.

\subsection{Ray Tracing}
Ray tracing is a model that enables a detailed representation of signal propagation and, in particular, multipath effects~\cite{tseng2017ray-tracing-assisted}.
This method simulates signal propagation in a specific environment by taking into account effects such as reflection, diffraction and scattering.
Ray tracing has been used since the 1990s to simulate the propagation of radio signals in urban environments~\cite{degli-esposti2014ray-tracing}.
Since then, ray tracing has become increasingly important in the planning and prediction of radio networks, as it offers deterministic modeling of propagation channels and can realistically reproduce the time and space dispersion characteristics of the channel that are important for modern wireless systems~\cite{degli-esposti2014ray-tracing}.

The basic idea of ray tracing is that electromagnetic waves are treated like rays and that their propagation can be calculated accurately~\cite{tseng2017ray-tracing-assisted}.
In a ray tracing simulation, the channel response between a transmitter and a receiver is calculated taking into account the geometric properties of the environment.
The simulation is carried out by tracing the rays that are emitted by the transmitter and that are either reflected, diffracted or scattered by the surrounding objects.
To do this, ray tracing requires a detailed description of the simulation environment, in particular the geometric and electromagnetic properties of the objects inside the environment~\cite{degli-esposti2014ray-tracing}.
The ray tracing approach thus enables accurate calculation of path losses and propagation delays and provides a precise prediction of the transmission channel~\cite{yun2015raytracing}.

A significant advantage of ray tracing over the statistical and stochastic models from network simulators is its ability to generate spatially consistent CIRs~\cite{hoydis2023learning}.
This ability is particularly needed for research topics in the field of wireless communication like 6G~\cite{hoydis2023sionnart}.
Although ray tracing provides precise results, the calculation of signal propagation requires a lot of computing power.
However, advancements in the computing power of modern computers, e.g., graphics processing unit (GPU), allow ray tracing to be carried out more efficiently today~\cite{degli-esposti2014ray-tracing}.


\subsection{Sionna}
%
%
Sionna is an open-source Python library based on tensorflow for link-level simulations~\cite{hoydis2023sionna}.
It has a ray-tracing extension called Sionna RT, which was developed for the realistic modeling of radio wave propagation. 
Sionna RT enables the simulation of specific radio environments which are described by a 3D model in which objects can be assigned with different materials having specific electromagnetic properties. 
These materials are characterized by parameters such as relative permittivity and conductivity, which allows the simulation of realistic interactions of radio waves with the surfaces~\cite{hoydis2023sionnart}.

Sionna RT calculates the paths of radio waves through the environment and tracks effects such as reflection, diffraction and scattering, enabling realistic modeling of the signal paths. 
This enables Sionna RT to calculate the CIR, taking into account environmental parameters such as antenna patterns or material properties.~\cite{hoydis2023sionnart}

Sionna RT is built on Mitsuba 3, a rendering system responsible for handling 3D scenes and calculating ray intersections within the 3D environment. 
This integration makes it possible to use Sionna RT for both indoor and outdoor radio wave simulation and provides a realistic and accurate modeling thanks to the detailed consideration of material properties and environmental structures.~\cite{hoydis2023sionnart}
Future developments of Sionna RT plan to integrate additional phenomena such as refraction to further increase the physical accuracy~\cite{hoydis2023sionnart}.
In summary, Sionna RT provides a tool for the simulation and modeling of radio wave propagation that enables the precise calculation of spatially consistent CIRs~\cite{hoydis2023sionnart}.

%

\subsection{Ns-3 Network Simulator}
The network simulator ns-3 is an event-driven, packet-based simulator developed primarily for research and academia in the field of communication networks.
It is an open-source software developed in C++ using object-oriented programming model~\cite{ns3}.
The broad research community continuously contributes to the expansion of the network simulator's functionality, which has led to a wide range of technologies (e.g., Ethernet, WiFi, LTE, WiMAX) and statistical and stochastic models for the simulation of channel attenuation, mobility and traffic generation~\cite{baidya2018flynetsim}.
ns-3 tries to reflect the reality as close as possible, therefore it uses several core concepts and abstractions that map well to how computer networks are built, i.e., a \texttt{Node} is a
fundamental entity connected to the network. It is a container for \texttt{Applications}, \texttt{Protocols} and \texttt{Network Devices}. An application is a user program that generates packet flows. A protocol represents a logic of network and transport level protocols. A \texttt{Network Device} is an entity connected to a \texttt{Channel} which represent the transmission medium between the devices.
For the simulation of Wi-Fi, either the \texttt{YansWifiChannel} or \texttt{SpectrumChannel} is used.
Each time a node sends a packet, the \texttt{YansWifiChannel} or \texttt{SpectrumChannel} calls a \texttt{PropagationLossModel} and \texttt{PropagationDelayModel} to calculate the path loss and transmission delay for the particular packet transmission. 
Also the movement of devices can be simulated by assigning mobility models to them that describe their movement in the simulation space.
Common mobility models in ns-3 are the \texttt{ConstantPositionMobilityModel} and the \texttt{RandomWalk2dMobilityModel}.
%
%
%
Finally, in ns-3, it is also possible to model buildings within the simulation in order to simulate both outdoor and indoor communication in the presence of buildings. 
For this purpose, ns-3 provides the Buildings module, which allows modeling buildings and provides additional propagation loss models that are specifically designed to calculate the propagation loss for outdoor and indoor transmissions.
Examples are \texttt{HybridBuildingsPropagationLossModel} and \texttt{ITUr1238PropagationLossModel}.

Another loss that is added to the total path loss is the shadowing loss. 
Therefore, in ns-3 a random loss value is generated for each pair of nodes and added to the total path loss. 
This value remains the same over the entire simulation run, and thus does not model any changes of the path loss over time. 
In addition, the shadowing losses are not spatially correlated as they are only generated using a log-normal distribution. 
This means that in a simulation with one transmitter and two very closely placed receivers, the shadowing losses generated for the transmissions to the receivers are completely different. 
The shadowing losses are also not symmetrical, which means that a different shadowing loss is used for a transmission from node a to node b than for a transmission from node b to node a.


%
%
%
\section{Motivation}\label{motivation}
The ns-3 provides a number of statistical and stochastic models for the simulation of signal propagation and the mobility of devices.
However, these models are not practical for the simulation of typical indoor scenarios as well as complex outdoor scenarios.
The propagation loss models in ns-3 either only considers a single reflection of the signal from the ground or uses probability distributions to model effects such as fading or shadowing. 
Moreover, the available mobility models like \texttt{RandomWalk2dMobilityModel} are unrealistic as the movement of a node can only be simulated within a simple bounding box. 
The same applies to the modeling of buildings in ns-3.
Buildings are only boxes and it is not possible to model rooms of different sizes or even doors. 
As not all rooms are rectangular in reality, these models can only be used in certain simplified scenarios.

To overcome these limits we developed \proposalName{} which introduces ray tracing into ns-3 as a way to provides a detailed and deterministic method for simulating radio wave propagation in complex 3D environments and enables the realistic representation of a wireless propagation channel.

\section{The \proposalName{} Framework}\label{design}
We designed an architecture that combines the ns-3 network simulator with the ray tracing functionality of Sionna RT.
This enables realistic simulation of the physical propagation of wireless signals within a given environment, taking into account important effects such as multipath propagation, fading and shadowing.

\subsection{Design Principles}\label{principles}
The integration of ray tracing functionality into a packet-level simulator like ns-3 is challenging.
The reason for this is the high computational effort required by ray tracing.
Table\,\ref{tab:exec_single_event} shows the average execution time on different hardware platforms for a single P2P channel computation in three different scenarios.
We can see differences of two orders of magnitude.
Moreover, the simulation of an outdoor channel requires the highest computational effort, i.e. the execution time is $\approx 5 \times$ larger in the selected outdoor vs. the simple indoor scenario.
Hence, already a 100-second simulation of such an outdoor scenario with a single WiFi cell, one AP and 32 STAs, would require the calculation of over 100 million P2P channels, resulting in an execution time of over 10 years.
Such a framework would make no sense from a practical point of view.
Finally, it is interesting to see that there is no advantage in using GPUs like the NVIDIA RTX 3060.
The calculation time is 3.5-6.5$\times$ slower compared to using a CPU with AMD Ryzen 5.
The reason for this is that we do not take the full advantage of the parallel execution capabilities of a GPU when we only calculate a single wireless P2P channel.

\begin{table}
    \centering
    \caption{Computation time for single CSI on different platforms: Ryzen9: Ryzen 9 7950X, Ryzen7: Ryzen 7 5800 and Ryzen7+RTX: Ryzen 7 5800 with RTX 3060 GPU}
    \label{tab:exec_single_event}
    \begin{tabular}{lccc}
        \toprule
                      & \multicolumn{3}{c}{Average P2P channel computation time [s]} \\
        Environment              & Ryzen9 & Ryzen7& Ryzen7+GPU\\
        \midrule
        Free-space     & 0.03 & 0.06 & 0.09\\
        Indoor (simple room)     & 0.9 & 1.2 & 7.7\\
        Outdoor (Munich)     & 4.54 & 5.32 & 18.46\\
        \bottomrule
    \end{tabular}
\end{table}

\noindent \proposalName{} targets the aforementioned challenge as follows:
\begin{enumerate}
    \item Exploitation of the \textbf{channel reciprocity}, i.e. the channel from $A$ to $B$ is the same as from $B$ to $A$,
    \item Exploitation of the \textbf{channel coherence time} $T_C$, i.e. channel recalculations can be avoided as for that period of time the wireless channel remains relatively stable,
    \item Avoiding computations towards \textbf{far-away receivers},
    \item Maximum parallelization through \textbf{predictive channel calculations}, e.g. usage of modern GPUs and multi-core CPUs.
\end{enumerate}

\begin{figure}
  \centering
  \includegraphics[width=0.9\linewidth]{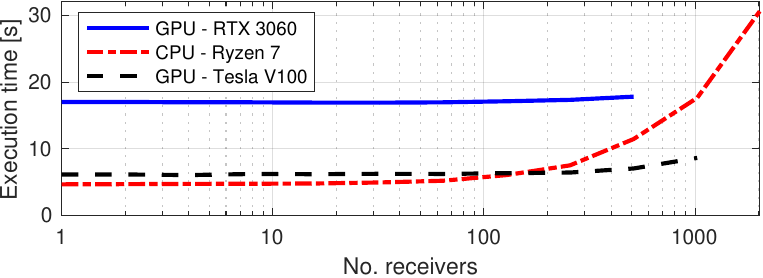}
  \vspace{-0.6em}
  \caption{Channel computation time in Sionna RT}
  \label{fig:benchmark_p2mp}
\end{figure}

\begin{figure}
  \centering
  \includegraphics[width=0.7\linewidth]{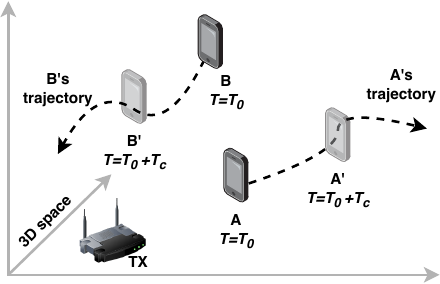}
  \vspace{-0.5em}
  \caption{To speed up execution \proposalName{} converts future node locations into locations of virtual nodes which can be calculated in parallel.}
  \label{fig:timespace}
  \vspace{-.8em}
\end{figure}

The first two points are addressed by caching channel state information (CSI) inside ns-3, i.e. for packet transmission within the coherence time $T_C$ the channel is not recomputed.
The second point is addressed as follows. 
First, the path loss is calculated using the Friis model in ns-3. 
If the received signal strength is below the noise level at the receiver, a calculation using ray tracing is omitted and this value is used instead.
Finally, the fourth point is addressed by the following observation.
Fig.\,\ref{fig:benchmark_p2mp} shows the channel computation time in pure Sionna for the Munich outdoor scenario with a single transmitter and a variable number of receivers $R$.
We can clearly see that even with a large $R$ the computation time is nearly constant, e.g. $R < 100$ for CPU and $R < 500$ for GPU (Tesla V100).
In case of GPUs the available video RAM limits the maximum size of $R$, e.g. 16\,GB are required for $R=1024$.
Hence, it is beneficial to avoid the calculation of each P2P channel individually and instead to offload the computation of the entire P2MP channel, i.e. from the sender to all potential receivers, to Sionna.
This approach leads to a large improvement, especially in dense networks.
However, even very low density networks, e.g. a scenario with only a single P2P link between an AP and STA, can benefit from parallelization.
The idea is to calculate in one step not only the current channel, but also the channels that might be needed in the future. 
This is possible because the mobility model, e.g. random walk, is independent of the actual simulated network traffic and thus the future position of the mobile nodes and their channels can be calculated in advance.
Fig.\,\ref{fig:timespace} illustrates the idea using an example of a single transmitter and two mobile receivers.
Here we convert the future two receiver node locations into the location of two additional virtual nodes.
Thereafter, the channel is computed towards all the four nodes
enabling its efficient parallel computation.
Note, a slight overhead is involved as currently the mobility model is computed sequentially for each mobile node.

In summary, we are leveraging two techniques, namely pre-caching of CSI and the inherent parallelization capabilities of Sionna, particularly when utilizing GPUs, resulting in significant reduction in execution time of simulations.

\subsection{Architecture}\label{arch}

\begin{figure}
  \centering
  \includegraphics[width=0.95\linewidth]{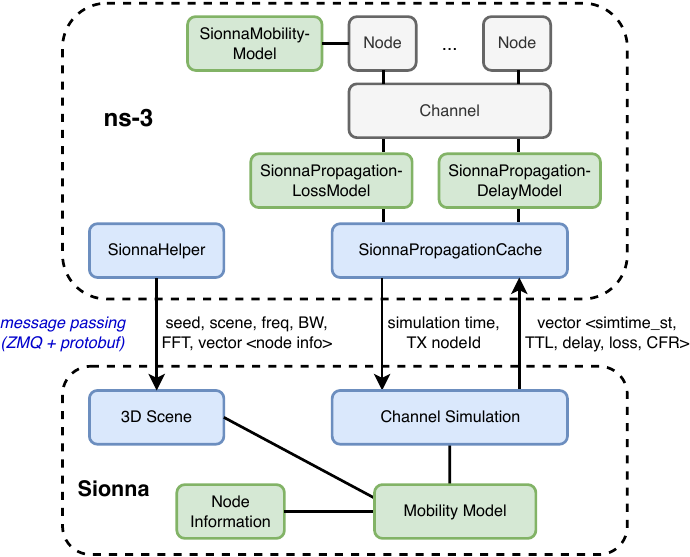}
  \caption{Architecture of the \proposalName{}}
  \label{fig:arch}
  \vspace{-.8em}
\end{figure}

The architecture of \proposalName{} is depicted in Fig.~\ref{fig:arch} and it shows the different components and their interaction.
We can identify two main components: Firstly, the ns-3 network simulator, which is extended with new models, and secondly, the channel simulation using Sionna RT. 
The key idea is to offload the calculation of the channel propagation of the wireless signal (CIR) from the transmitter to the receiver for each packet transmission in ns-3 to Sionna RT and then return realistic information about the path loss, channel state information and propagation delay to ns-3.
Hence, the two new models in ns-3, namely
\texttt{SionnaPropagationLossModel} and \texttt{SionnaPropagationDelayModel} are wrappers for Sionna RT.
Moreover, the simulation of mobility is offloaded from ns-3 to Sionna RT using \texttt{SionnaMobilityModel}.
This is because the existing 2D random movement model in ns-3 only allows the random movement of nodes within a rectangular bounding box.
In \proposalName{} the nodes can move within the more complex boundaries of the 3D model.

The \texttt{SionnaHelper} class is responsible for the life-cycle management.
At the start of the simulation all relevant information like the name of the XML file of the scene to be simulated, the center frequency, channel bandwidth, FFT size (i.e., number of OFDM subcarriers), used mobility model as well as general information about each node are passed from ns-3 to Sionna RT.
During the simulation new channel propagation information between a pair of wireless nodes, i.e. sender and receiver, are requested on-demand by ns-3.
In order to benefit from parallelization of the ray tracing process, the requested P2P channel is extended towards the computation of the full P2MP channel, i.e. from the sender to all potential receiver nodes.
Moreover, depending on the available resources, e.g. video memory of GPU, not only the current channel, but also the channels that might be needed in the future are computed by simulating mobility and placing additional virtual nodes (Fig.\,\ref{fig:timespace}).
%
%
Then, using Sionna’s ray tracing capabilities, the propagation paths are calculated, including both LOS
and reflected paths.
From the paths the CIR and afterwards the \ac{CFR} is calculated.
The propagation loss is calculated from the root mean square of the \ac{CFR}, while the propagation delay is calculated from the minimum of the delays of all paths. 
The propagation delay thus corresponds to the delay of the shortest path between the transmitter and receiver.
In addition, the \ac{TTL} is calculated using Eq.\,\ref{eq:coherence_time_fundamentals} which is used by caching, i.e., for packet transmission within the coherence time $T_C$ the channel is not recomputed.
%
%
The two components, ns-3 and Sionna, are synchronized using the simulation time provided by ns-3 in each request.
In order to improve the execution speed of the framework all channel state data, including the speculative calculation of future channels, is cached inside \texttt{SionnaPropagationCache} for the duration of $T_C$.
Moreover, the cache accounts for channel reciprocity, i.e., channel state information calculated for a transmission from node $A$ to node $B$ can also be used for the opposite transmission from node $B$ to node $A$.


%

%
In order to have more realistic mobility in 3D environments also the mobility model was offloaded to Sionna RT.
This design decision was needed to ensure that mobile nodes do not move through walls or other obstacles in complex 3D environments, but bounce off them correctly.
Moreover, this enables the implementation of channel pre-caching, i.e. the speculative calculation of channels which might be needed in the future.
For the mobility model ray tracing is used where we shoot a ray from the current node position in the walking direction to calculate when the node hits an object and how it bounces off it.
Note, that the modeling of movement of nodes using ray tracing results in an increased execution time.

%

\subsection{Implementation}
To enable the interactions between the two software components we selected the client-server model.
Here the ns-3 simulator acts as client while the Sionna process acts as a server.
For inter-process communication (IPC) we selected message passing as it gives are the possibility to run ns-3 and Sionna on different computers.
The latter could be equipped with GPUs to accelerate the raytracing computations within Sionna RT.
This also simplified the interaction between the two frameworks as both use different programming language, i.e., C++ vs. Python.
As messaging library we selected ZeroMQ in request-response pattern that allows to send channel state requests from ns-3 to Sionna and return responses. 
Importantly, it enables synchronous communication, which ensures the discrete execution of the ns-3 simulation. 
For the serialization of the messages we used protocol buffers to ensure efficient transmissions.
Finally, the random mobility model we implemented using the Python library Mitsuba.

%
%
%
%

Our \proposalName{} software package together with clarifying examples is provided to the community as open-source under a GPL in our online repository: \mbox{\texttt{https://github.com/tkn-tub/ns3sionna}}. 



%
\section{Validation}

\begin{figure}
  \centering
  \includegraphics[width=\linewidth]{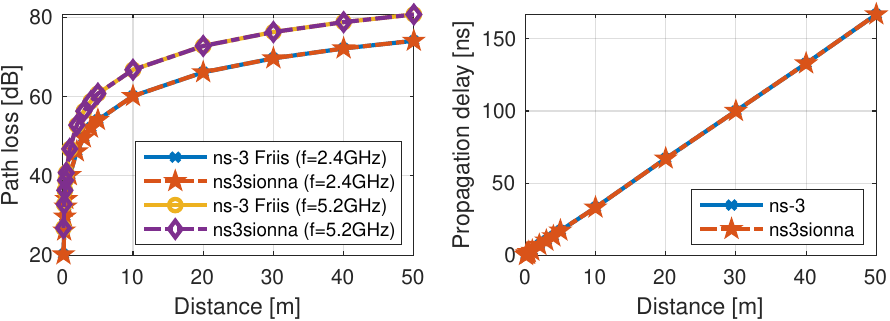}
  \vspace{-0.5em}
  \caption{Validation of path loss and propagation delay in free-space}
  \label{fig:free_space_valid}
  \vspace{-.8em}
\end{figure}

For validation we selected a free-space environment without any walls or obstacles.
Here we compared the signal pathloss computed in \proposalName{} with the results obtained in pure ns-3 using the \texttt{FriisPropagationLossModel} model.
For delay we used \texttt{ConstantSpeedPropagationDelayModel} which calculates the propagation delay from the distance $d$ between the transmitter and receiver and the signal propagation speed $c$, i.e., speed of light, as $\tau = d / s$.
As can be seen from Fig.\,\ref{fig:free_space_valid} the results are the same.


%
%
%
\section{Example Scenarios}\label{sec:scenarios}
In this section we present results for two scenarios, indoor and outdoor, using \proposalName{}.
\subsection{Indoor Scenario}
\begin{figure}
  \centering
  \includegraphics[width=0.45\linewidth]{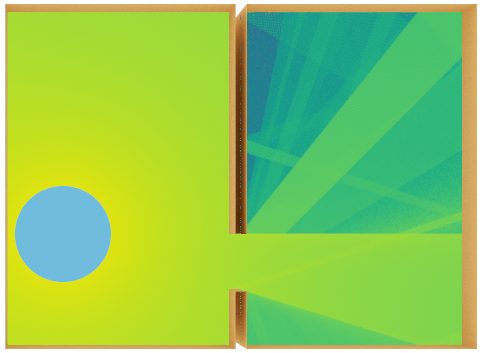}
  \caption{Indoor scenario - two rooms and open door}
  \label{fig:2_rooms_with_door_open}
\end{figure}

\begin{figure}
  \centering
  \includegraphics[width=\linewidth]{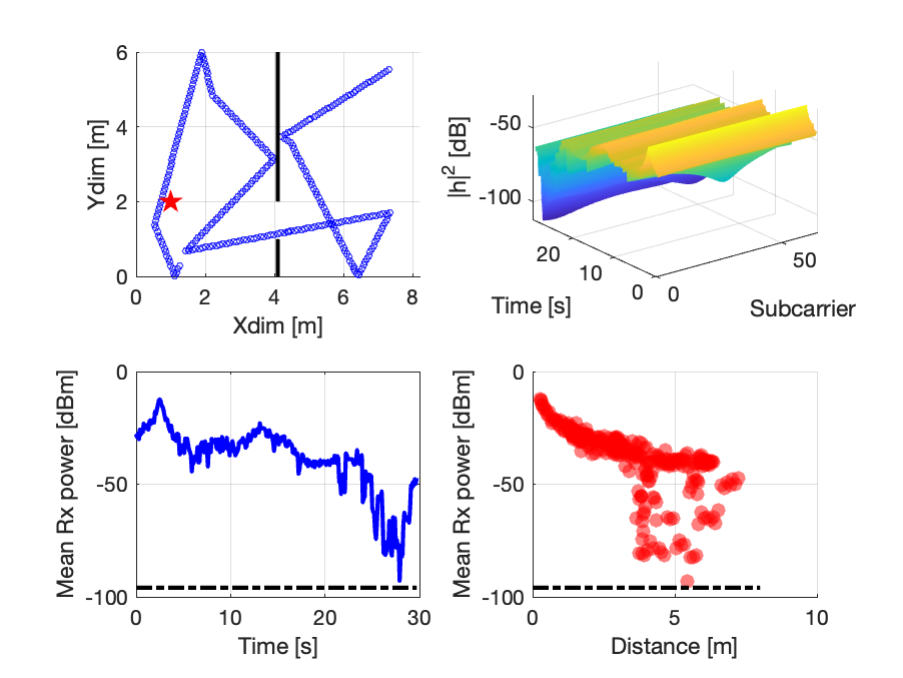}
  \vspace{-1.5em}
  \caption{Results for indoor scenario}
  \label{fig:ex1_paper}
  \vspace{-0.8em}
\end{figure}

The indoor scenario consists of two adjacent rooms connected with an open door (Fig.\,\ref{fig:2_rooms_with_door_open}).
In this environment, a WiFi signal (802.11ac, 5\,GHz, 20\,MHz channel) sent from the \ac{AP} in the left room must therefore pass through the doorway and then must be reflected by one or more walls in the second room in order to a reach a \ac{NLoS} STA located in the other room. 
As a result, the signal sometimes has to travel a very long distance and loses a lot of energy accordingly.

The results using the random walk mobility model ($v=1\,m/s$) are shown in Fig.\,\ref{fig:ex1_paper} where the upper left figure shows the location of the AP (red star) and the trajectory of the STA (blue).
The resulting receive power $P_{\mathrm{rx}}$ is shown in lower left figure.
Here we can clearly see both the spatial and the time correlation of $P_{\mathrm{rx}}$, i.e., smooth changes in both space and time.
The impact of distance between AP and STA on $P_{\mathrm{rx}}$ is shown in lower right figure.
We can clearly see the impact of the wall separating the two rooms, i.e., variation in $P_{\mathrm{rx}}$ by up to 50\,dB for nearly the same distance. 
Finally, the upper right figure shows the magnitude of the CFR where we can observe some frequency-selectivity.

\subsection{Outdoor Scenario}
\begin{figure}
  \centering
  \includegraphics[width=0.5\linewidth]{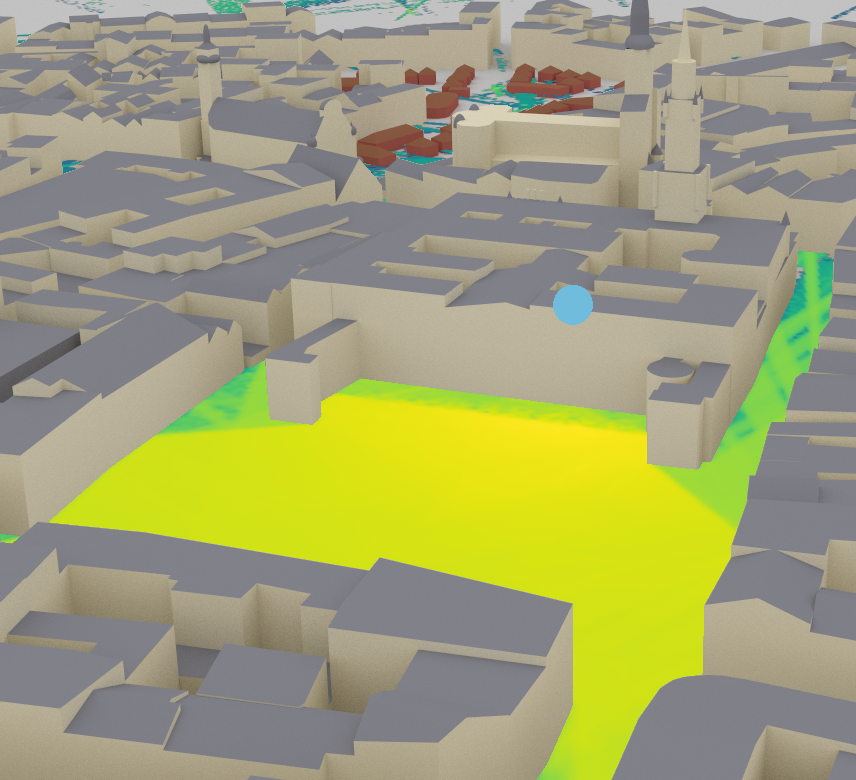}
  \caption{Outdoor scenario: area around Frauenkirche in Munich}
  \label{fig:munich_scenario}
\end{figure}

For the outdoor environment we selected the example scene from Sionna containing the area around the Frauenkirche in Munich (Fig.\,\ref{fig:munich_scenario}).
Again 802.11ac at 5\,GHz was used but the range of motion and STA speed ($v=7\,m/s$) was larger than in the indoor scenario.

\begin{figure}
  \centering
  \includegraphics[width=\linewidth]{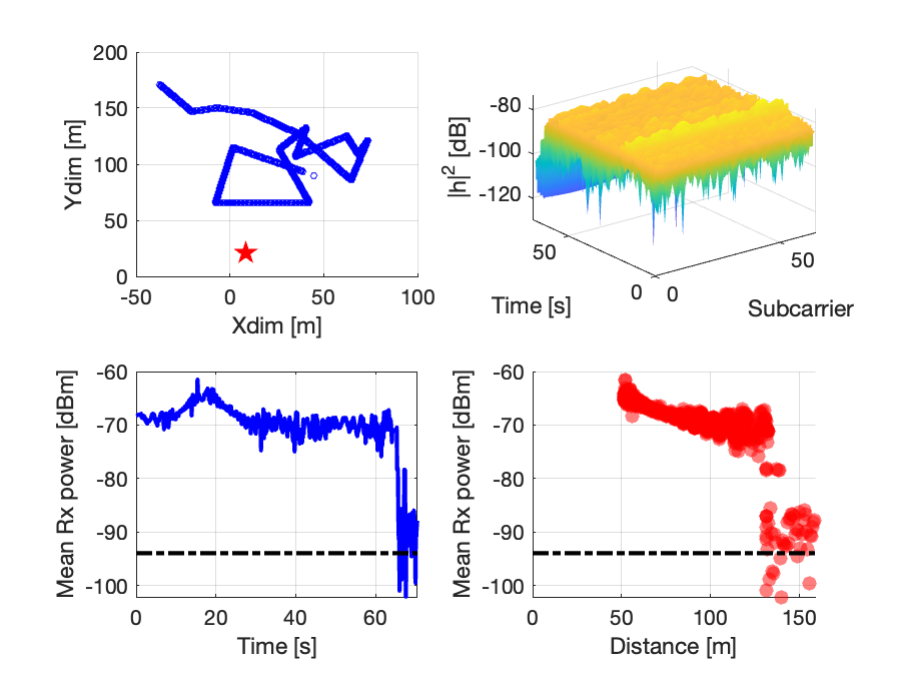}
  \vspace{-1.5em}
  \caption{Results for outdoor scenario}
  \label{fig:ex2_munich_paper}
  \vspace{-.8em}
\end{figure}

From the magnitude of the CFR in Fig.\,\ref{fig:ex2_munich_paper} we can observe a highly-frequency selective channel.
Moreover, up-to a distance of approx. 125\,m the $P_{\mathrm{rx}}$ does not change much due to the dominant LOS component.
Beyond that distance $P_{\mathrm{rx}}$ drops quickly due to the NLoS condition.

\section{How to use \proposalName{}?}\label{sec:howto}

\subsection{Adjustments in ns-3 Script}
Any existing ns-3 script can be adapted to be used with \proposalName{}.
The WiFi module is currently supported, with support for LTE planned for the future.
An example script is shown in Listing\,\ref{list:exns3}.
Using \texttt{SionnaHelper} (line 2) we configure the framework for the Munich outdoor scenario to run on the same host, i.e. both the ns-3 and Sionna process are executed on same machine.
In the following lines the cache, delay and loss models from \proposalName{} configured and are connected to the \texttt{YansWifiChannel}.
Later the used signal frequency and channel bandwidth are set (line 17-19).
Finally, the \texttt{SionnaHelper} is started.
Note, no changes are needed on the server-side component of \proposalName{} as all information is passed via ZMQ at runtime.

\lstset{language=Python,
        basicstyle=\ttfamily\scriptsize,
        columns=fullflexible,
        xleftmargin=7pt,
        xrightmargin=7pt,
        captionpos=b, 
        frame=single,
        framexleftmargin=0pt,
        framexrightmargin=0pt,
        breaklines=true,
        keepspaces=true,
        numbers=left,
        numbersep=5pt,
        numberstyle=\tiny\color{gray},
        showlines=true,
        postbreak=\mbox{\textcolor{red}{$\hookrightarrow$}},
        showstringspaces=false,
        keywordstyle=\color{blue},
        commentstyle=\color{comments},
		emphstyle=\ttb\color{deepred},
        }

\begin{lstlisting}[language=c++, caption=Example ns-3 script showing how to interact with \texttt{ns3sionna}, label=list:exns3]
SionnaHelper sionnaHelper("munich/munich.xml", "tcp://localhost:5555");

Ptr<YansWifiChannel> channel = CreateObject<YansWifiChannel>();
Ptr<SionnaPropagationCache> propagationCache = CreateObject<SionnaPropagationCache>();
propagationCache->SetSionnaHelper(sionnaHelper);
Ptr<SionnaPropagationDelayModel> delayModel = CreateObject<SionnaPropagationDelayModel>();
delayModel->SetPropagationCache(propagationCache);
Ptr<SionnaPropagationLossModel> lossModel = CreateObject<SionnaPropagationLossModel>();
lossModel->SetPropagationCache(propagationCache);
channel->SetPropagationLossModel(lossModel);
channel->SetPropagationDelayModel(delayModel);

sionnaHelper.Configure(get_center_freq(apDevices.Get(0)), get_channel_width(apDevices.Get(0)));
sionnaHelper.Start();
Simulator::Run(); 
Simulator::Destroy();
\end{lstlisting}
\vspace{-5pt}

\subsection{Usage of Custom 3D Scenes}
For a realistic simulation a detailed 3D model of the environment is needed where objects are assigned different materials that describe their electromagnetic properties.
Some example scenes (Fig.\,\ref{fig:2_rooms_with_door_open}) are provided in \proposalName{} but the user can add its own.
Custom 3D scenes can be created using the free and open source 3D graphics software Blender\,\cite{blender}. 
As Sionna imports the scenes in the form of an XML file, the following workflow is used: First, the 3D model is designed in Blender. 
Then the individual objects, such as the floor, walls and ceiling, are exported as separate OBJ files containing the geometric data of the objects. 
Finally, an XML file is created in which the objects from the OBJ files were linked to their corresponding materials.

\section{Benchmarking}\label{sec:eval}
Channel simulation using ray tracing has a high computational effort.
Therefore, in this section we want to benchmark \proposalName{} and compare its performance with pure ns-3 in terms of execution time.
Specifically, we want to investigate which parameters have a decisive influence on the execution speed of our framework.
For benchmarking we run an indoor scenario (single room) with a configuration using 802.11ac WiFi (20\,MHz channel) with a single AP and variable number of STAs $R$.
The following three scenarios were selected with different downlink UDP packet rates ($U$) and STA speed ($v$):
\begin{itemize}
    \item \textbf{hT/hM} - high traffic (U=50\,p/s per STA), high mobility (v=7\,m/s, $T_C$ = 1.6-3.2\,ms)
    \item \textbf{lT/lM} - low traffic (U=1\,p/s per STA), low mobility (v=1\,m/s, $T_C$ = 11.1-22.3\,ms)
    \item \textbf{hT/zM} - high traffic (U=50\,p/s per STA), stationary
\end{itemize}

\subsection{Usage of CPU}
%

\begin{figure}
  \centering
  \includegraphics[width=\linewidth]{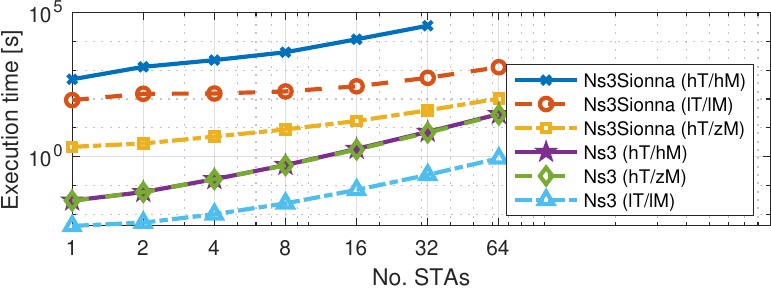}
  \vspace{-1.5em}
  \caption{Framework execution time: ns-3 vs. \proposalName{}}
  \label{fig:benchmark_hpc9_cpu}
\end{figure}

\begin{figure}
  \centering
  \includegraphics[width=\linewidth]{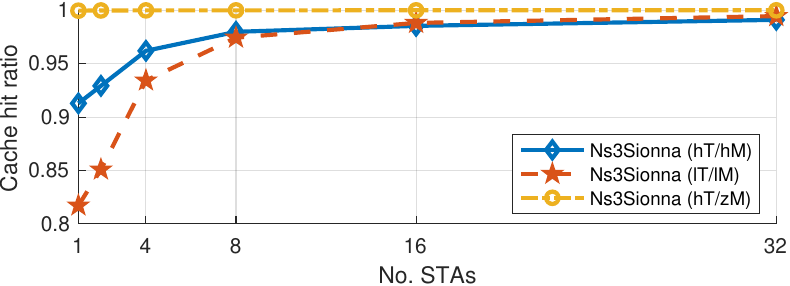}
  \vspace{-1.5em}
  \caption{Caching performance of \proposalName{}}
  \label{fig:caching_hpc9_cpu}
  \vspace{-0.8em}
\end{figure}

Fig.\,\ref{fig:benchmark_hpc9_cpu} shows the execution time on a host with AMD Ryzen 9 7950X (128\,GByte RAM, 16 cores) to perform a single simulation run of 10\,s.
First, we can clearly see the increase in execution time vs. pure ns-3.
The increase is smallest in the stationary scenario (\textbf{hT/zM}) as the channel between each pair of nodes need only be computed once resulting in an increase of execution time for \proposalName{} at around 3.6-73 $\times$ vs. pure ns-3.
The performance loss becomes smaller for larger network sizes because \proposalName{} is able to efficiently compute the P2MP channel in one execution step in Sionna.
Moreover, the predictive channel calculation helps in execution time reduction which is visible from the increasing cache hit rate for larger network sizes (Fig.\,\ref{fig:caching_hpc9_cpu}).
Finally, due to higher network traffic more calculations have to be performed in ns-3.

The impact of ray tracing is much larger in the two mobile scenarios.
Here the channel must be calculated very frequently, e.g. 625\,Hz in \textbf{hT/hM}, resulting in up-to four orders of magnitude larger execution time.
However, we also see that with the increasing network size the performance loss over pure ns-3 sharply decreases.
In the \textbf{lT/lM} scenario with 64 STAs the execution time of \proposalName{} is $1400\times$ larger vs. pure ns-3.
The difference is similar in the \textbf{hT/hM} scenario, i.e. with 16 STAs the execution time is $6200\times$ larger. 

Finally, the selected scene (indoor or outdoor) has a large impact on the execution time as the raytracing requires more computation in large environments with lots of objects.
Table.\,\ref{tab:exec_single_event} shows the average execution time on different hardware platforms for a single channel computation in three different scenarios.
The execution time is $\approx 5 \times$ larger in the selected outdoor vs. the simple indoor scenario.
This is the reason why \proposalName{} uses pre-caching to avoid unnecessary expensive channel calculations.
%
%

%
\subsection{Usage of GPU}
\begin{figure}
  \centering
  \includegraphics[width=\linewidth]{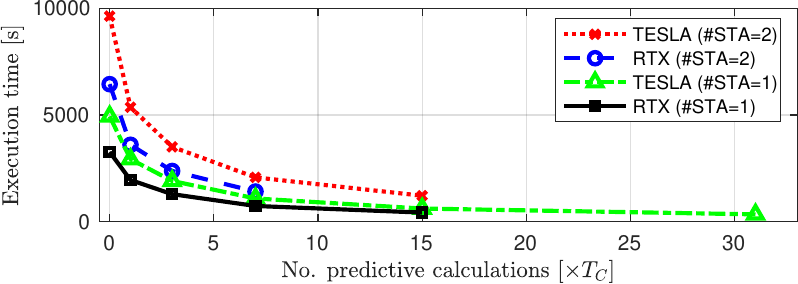}
  \vspace{-1.5em}
  \caption{Speedup through predictive calculations}
  \label{fig:eval_cmp_gpus}
  \vspace{-0.8em}
\end{figure}

Fig.\,\ref{fig:eval_cmp_gpus} shows the speedup in execution time when performing predictive channel calculations for two different GPUs, Nvidia Tesla V100 (16GB) and Nvidia RTX 3060 (8GB).
The improvement is the highest for the Tesla GPU with a factor of $14.2548\times$ in a network with single STA.
With larger network size the speedup decreases to a factor of $7.9023\times$ as the maximum number of predictive calculations is limited by the available GPU memory size.
With the RTX GPU having only half the memory size of the Tesla the speedup is $7.6\times$ and $4.6\times$ for a network with one and two STAs respectively.
In summary, the results clearly show that \proposalName{} can be scaled with the GPU memory size.

%
%
\subsection{Discussion}\label{discussion}
The proposed \proposalName{} framework has limitations which we want to address in the future.
First, the simulation of large 3D scenarios with dense high mobile, high traffic nodes is challenging due to the very high execution time.
The usage of modern multi-core CPUs and GPU cards with large video memory are recommended and it is already supported by Sionna and hence \proposalName{}.
%
%
%
Additionally, parallelization of the ray tracing process is feasible, as the computations for individual P2MP channels are independent and can be executed in parallel across multiple hosts or GPUs.
Second, so far our framework supports only SISO, i.e., single antenna APs and STAs.
However, it is straightforward to extend it to MIMO as Sionna already supports different kind of antenna types.
Nevertheless, suitable PHZ abstraction models for MIMO must be found to keep the computational effort low.
Third, we focused ourself on the support of WiFi.
However, \proposalName{} is not limited to WiFi and can be used together with the \texttt{SpectrumChannel} from ns-3.
This allows the usage of other modules like LTE.
Finally, as Sionna supports Reconfigurable Intelligent Surfaces (RIS), our framework can be extended to enable simulation of RIS inside ns-3.
This would create the opportunity to study the impact of RIS on higher network protocol layers (e.g., MAC layer).


%
%
%
\section{Related Work} \label{chapter:related_work}
%

The related work covers both the use of ray tracing methods for the simulation of wireless channel propagation as well as relevant extensions of network simulators.

Yun et al.\,\cite{yun2015raytracing} reviews the basic concepts for radio propagation modeling using ray tracing methods.
Sionna\,\cite{hoydis2023sionna} is an open-source library for simulating the physical layer of wireless and optical communication systems.
\proposalName{} extends Sionna with the ability to perform system-level simulations by providing the higher layers of the network protocol stack, i.e. data link, network, transport and application layer.
\textcite{dricot2004high} developed a realistic physical model for the ns-2 simulator to overcome the limitations of traditional statistical models, particularly in indoor environments. Their approach integrates ray tracing with Markov chains: ray tracing is employed to compute the mean propagation loss by accounting for signal reflections and diffractions caused by objects in the environment, while Markov chains model the time-varying effects of signal fading. Evaluation results show an increase in execution time by a factor of 100 as compared to conventional ns-2 models, primarily due to the computational complexity of ray tracing and Markov chain calculations.
\textcite{degli-esposti2014ray-tracing} explored various beamforming strategies and proposed a novel approach utilizing ray tracing to identify and simulate the strongest signal paths between a transmitter and receiver. This information was then used to direct the beams along these optimal paths. Their method demonstrated a significant improvement in signal quality at the receiver, highlighting its effectiveness compared to traditional radial beamforming techniques.
\textcite{seretis2022hybrid} introduced a machine learning-based approach to predict received signal strength (RSS) in indoor environments, addressing the challenge of limited availability of suitable training data. They proposed that when measurement data is insufficient for accurate modeling, the inclusion of synthetically generated data can enhance RSS prediction performance. To this end, the authors developed a hybrid machine learning model trained on both measured and synthetically generated data, with the synthetic data produced using ray tracing. Evaluation results demonstrated that the hybrid model achieved significantly higher accuracy in RSS prediction compared to models trained exclusively on either measured or simulated data.




%
%
%
\section{Conclusion}\label{sec:conclusion}
We presented \proposalName{}, a software module that brings realistic channel simulation using ray tracing to the widely used ns-3 network simulator.
Our plans for the future are focused on performance improvements by further parallelizing the channel computations through distributing the load over multiple GPU/CPU nodes.
Moreover, we envision to extend our framework to enable the support of MIMO and Reconfigurable Intelligent Surfaces (RIS) with proper PHY abstractions models enabling the research of next-generation WiFi protocols.
\section*{Acknowledgements}
This work was supported by the Federal Ministry of Education and Research (BMBF, Germany) within the 6G Research and Innovation Cluster 6G-RIC under Grant 16KISK020K as well as by the German Research Foundation (DFG) within the project ML4WiFi under grant DR 639/28-1.


\printbibliography

\end{document}